\documentclass[12pt]{article}
\usepackage[dvips]{graphicx}
\usepackage[usenames]{color}
\begin{document}
\begin{center}
{\LARGE\bf Exact Analysis of Scaling and Dominant Attractors Beyond
the Exponential Potential}
 \vskip 0.15 in
 $^\dag$Wei Fang$^{1, 2}$, Ying Li$^4$, Kai Zhang$^3$, Hui-Qing Lu$^3$\\
\small {$^1$\cal$ Department ~of ~Physics, ~Shanghai~ Normal~ University, ~Shanghai, ~200234, ~P.R.China$\\
$^2$\cal$ The ~Shanghai~ Key ~Lab ~of ~Astrophysics, ~Shanghai, ~200234, ~P.R.China$\\
$^3$\cal$ Department~of~Physics,~Shanghai~University, ~Shanghai,~200444, ~P.R.China$\\
$^4$\cal$ College~ of ~Information~ Technology, ~Shanghai~ Ocean~
University, ~Shanghai,~ 201306, ~P.R.China$}
 \footnotetext{$\dag$wfang@shnu.edu.cn}
 \vskip 0.5
in  \centerline{\bf Abstract} \vskip 0.2 in
\begin{minipage} {5.8in} {\hspace*{10pt}\small By considering the potential parameter
$\Gamma$ as a function of another potential parameter $\lambda$[47],
We successfully extend the analysis of two-dimensional autonomous
dynamical system of quintessence scalar field model to the analysis
of three-dimension, which makes us be able to research the critical
points of a large number of potentials beyond the exponential
potential exactly. We find that there are ten critical points in
all, three points $P_{3, 5, 6}$} are general points which are
possessed by all quintessence models regardless of the form of
potentials and the rest points are closely connected to the concrete
potentials. It is quite surprising that, apart from the exponential
potential, there are a large number of potentials which can give the
scaling solution when the function $f(\lambda)(=\Gamma(\lambda)-1)$
equals zero for one or some values of $\lambda_{*}$ and if the
parameter $\lambda_{*}$ also satisfies the condition Eq.(16) or
Eq.(17) at the same time.  We give the differential equations to
derive these potentials $V(\phi)$ from $f(\lambda)$. We also find
that, if some conditions are satisfied, the de-Sitter-like dominant
point $P_4$ and the scaling solution point $P_9$( or $P_{10}$) can
be stable simultaneously but $P_9$ and $P_{10}$ can not be stable
simultaneity. Although we survey scaling solutions beyond the
exponential potential for ordinary quintessence models in standard
general relativity, this method can be applied to other extensively
scaling solution models studied in literature[46] including coupled
quintessence, (coupled-)phantom scalar field, k-essence and even
beyond the general relativity case $H^2 \propto\rho_T^n$.
 we also discuss the disadvantage of our
approach.
\\
 {\hspace*{15pt}\small 
}
\end{minipage}
\end{center}

\newpage
\section{Introduction}
 Scalar fields had played an essential role in modern cosmology in the past semi-century. This assumed scalar field had been used
 for various purposes in different cosmological research aspects[1], such as to drive inflation, to explain a time variable cosmological "constant"
 and so on. Especially, after the discovery of the
 accelerating expansion of universe, it has gained another hotly discussion as the candidate for dark energy. There are so many
 scalar field dark energy models, such as quintessence model[2-13], non-canonical scalar field model
 ( including K-essence[14-17], phantom[18-22], B-I scalar field[23-29] and so on) and coupled scalar field model[30-31].
 There are also detailed studies on the multi-scalar field models which  give an
  effective state equation $w_{eff}$ passing through the phantom divide line ($w=-1$)[32-34]. Some of these multi-scalar
  field models[35-36] can always evolve onto the regime of scalar field dominance ${\lambda_{eff}}^2>3 \gamma$ even
  if each field has too steep a potential to drive the accelerating expansion. For all of these scalar field models we mention above, the important
  thing is to choose different form of kinetic terms and different potentials from a fundamental physical motivation
  or directly from the observation. As are expected, these different scalar field models will give different cosmological evolutions,
 different evolutions of state equation $w$, different values of sound speed $c_s^2$ and different cosmological perturbation.
 So they can in principle be distinguished or excluded by the increasing observation data.
 \par The phase-plane analysis of the cosmological autonomous  system is an effective method to find the cosmological
  scaling and dominant attractor solutions. A phase-plane analysis of
  cosmologies containing a barotropic fluid and a scalar field with an exponential potential was
  presented[37].  Hao and Li studied the attractor solution of phantom scalar field with the exponential
  potential[38-39]. On the other hand, L.Amendola considered the
  case of coupled quintessence[31].
  The case of phantom scalar field interacting with dark matter was also
  investigated[40-41]. Guo also investigated the properties of the critical points of
multi-field model with an exponential potential[42-43] One may
realize that the potentials investigated in all these papers are the
exponential form.
 Disregarding the important roles of the exponential potential in higher-order or higher-dimensional
 gravity theories and string or kaluza-klein type models,
 the reason that why they are choosing the exponential potential may be that, only the exponential potential can give a
 two-dimension autonomous system. Since in this case the value of the parameter $\Gamma$ equals 1 and then
  another parameter $\lambda$ equals a constant(see Eq.(4) for the definition of parameters $\Gamma, \lambda$ ),
   so the system(see Eqs.(5-7) below) will reduce to the
 two-dimension autonomous system. However, authors also considered the more complicated case
 when $\lambda$ is a dynamically changing quantity[44-46].
They applied the discussion of constant $\lambda$ to this case and
obtained the so-called "instantaneous" critical points. For example,
if $\Gamma$ is a constant(but does not equal one), saying
$\Gamma=(n+1)/n$, the corresponding potential is the inverse
power-law potential $V(\phi)=V_0{\phi}^{-n}$ with $n>0$. One of the
critical points $(x_c, y_c)=(\lambda/\sqrt{6},
[1-\lambda^2/6]^{1/2})$ will become the "instantaneous" critical
point $(x(N)=\lambda(N)/\sqrt{6}, y(N)=[1-\lambda(N)^2/6]^{1/2})$.
When $\Gamma>0$, $\lambda(N)$ will decrease toward zero, then the
"instantaneous" critical points will eventually approach $x(N)
\rightarrow 0$ and $y(N) \rightarrow 1$. This method is not exact
here and obviously the critical point is not a true critical point.
Recently, a solution of multiple-attractor in three-dimension
autonomous system of the quintessential models was studied in
literature[47]. After writing the parameter $\Gamma$ as a function
of $\lambda$, the author obtained a tracker solution which is
different from those discovered before and found a solution of
multiple-attractor. Here we will extend the idea to an arbitrary
function $\Gamma(\lambda)$. We will find out all the critical points
of the dynamical autonomous system, and then investigate the
properties of the critical points and their cosmological
implications in general. Regarding parameter $\Gamma$ as a function
of $\lambda$ is a quiet efficient approach since we can investigate
many quintessence models with different potentials. Giving a
concrete form of function $\Gamma(\lambda)$ is equivalent to give a
concrete form of potential $V(\phi)$ since we can in principle
figure out the potential via the relation between parameter $\Gamma$
and $\lambda$. What are the general properties of the critical
points when we consider the three-dimension autonomous system? Does
there also exists scaling solution when we consider any function of
$\Gamma(\lambda)$? Among all the critical points which critical
points are the critical points for all quintessence and which are
only relative to the concrete potentials? In our paper, we will try
to shed light on these issues. The paper is organized as follows: in
Section 2 we present the theoretical framework and give the
differential relation between the function $\Gamma(\lambda)$ and
potential $V(\phi)$. We find out all the critical points and
investigate their properties in Section 3. We try to give the
cosmological implications of these critical points in section 4. We
briefly display our conclusions in section 5.
\section{Basic theoretical frame} We start with a spatially flat Friedman-Robertson-Walker universe containing a scalar field $\phi$ and a barotropic fluid (with state equation
$p_b=w_b\rho_b$). To simply, we give the Einstein equations
directly:
\begin{equation}H^2=\frac{\kappa^2}{3}[\frac{1}{2}\dot{\phi}^2+V(\phi)+\rho_b]\end{equation}
\begin{equation}\dot{H}=-\frac{\kappa^2}{2}[\dot\phi^2+(1+w_b)\rho_b]\end{equation}
The motion equation of the scalar field $\phi$ is:
\begin{equation}\ddot{\phi}+3H\dot{\phi}+\frac{dV}{d\phi}=0\end{equation}
Following[48], we define the following dimensionless variables:
\begin{equation}x=\frac{\kappa\dot\phi}{\sqrt6 H}, y=\frac{\kappa\sqrt{V}}{\sqrt3 H},
 \lambda=-\frac{V'}{V}, \Gamma=\frac{V V''}{V'^2}\end{equation}
Where $V'=dV(\phi)/d\phi, V''=d^2V(\phi)/d\phi^2 $.
 Using Eq.(4), Eqs.(1-3) can be rewritten in the following dynamical form[37, 46, 48]:
\begin{equation}\frac{dx}{dN}=-3x+\frac{\sqrt{6}}{2}\lambda y^2+\frac{3}{2}x[(1-w_b)x^2+(1+w_b)(1-y^2)]\end{equation}
\begin{equation}\frac{dy}{dN}=-\frac{\sqrt{6}}{2}\lambda xy+\frac{3}{2}y[(1-w_b)x^2+(1+w_b)(1-y^2)]\end{equation}
\begin{equation}\frac{d\lambda}{dN}=-\sqrt{6} \lambda^2(\Gamma-1)x\end{equation}
where $N=ln(a)$. Here we should emphasize that Eqs.(5-7) is not a
dynamical autonomous system since the parameter $\Gamma$ is unknown.
However, if we consider $\Gamma$ as a function of $\lambda$, namely
\begin{equation}\Gamma(\lambda)=f(\lambda)+1\end{equation}
then Eq.(7) becomes:
\begin{equation}\frac{d\lambda}{dN}=-\sqrt{6}\lambda^2f(\lambda)x\end{equation}
Hereafter, Eqs(5-6) and Eq.(9) are definitely a dynamical autonomous
system. We will see that $\Gamma$ as a function of $\lambda$ can
cover many quintessential potentials. The three-dimension autonomous
system reduces to two-dimension autonomous systems when
$f(\lambda)=0$. In this case, the potential is the exponential form
which has been completely studied in many literatures. When
$f(\lambda)$ equals a nonvanishing constant $f_{\lambda}$, then the
potential is proportional to $(c_1\phi+c_2)^{-1/f_{\lambda}}$, which
is just the potential which has been considered as "instantaneous"
critical points[48]. Generally speaking, we can analyze any explicit
function. For some more complicated form,
$\Gamma(\lambda)=1+\frac{1}{n}-\frac{n\sigma^2}{\lambda^2}$
corresponds to $V(\phi)=\frac{V_0}{[cosh(\sigma\phi)]^n}$,
$\Gamma(\lambda)=1 \pm \frac{\alpha}{\lambda^2}$ corresponds to
$V(\phi)=V_0e^{\pm \alpha \phi(\phi+\beta)/2}$,
$\Gamma(\lambda)=1+\frac{2}{\sqrt{\lambda}}$ corresponds to
$V(\phi)=V_0e^{1/\phi}$. The form of
$\Gamma(\lambda)=1+\frac{1}{\beta}+\frac{\alpha}{\lambda}$, which
corresponds to $V(\phi)=\frac{V_0}{(\eta+e^{-\alpha\phi})^{\beta}}$,
was considered as an interesting cosmological model where the
universe can evolve from a scaling attractor to a de-Sitter-like
attractor by introducing a possible mechanism of spontaneous
symmetry breaking[47].
\par In the paper[47], the author
gave an approach to obtain the potential $V(\phi)$ as follows: Since
the potential $V(\phi)$ is only a function of the field $\phi$, then
the parameters $\lambda$ and $\Gamma$ can be written as a function
of field: $\lambda=P(\phi), \Gamma=Q(\phi)$. If the inverse function
of $P(\phi)$ exists, then we have:
\begin{equation}\Gamma=Q(P^{-1}(\lambda))\equiv {\cal F}(\lambda)\end{equation}
Using the definition of $\lambda$ and $\Gamma$, $V''$ can be written
as $V''=\frac{V'^2}{V}{\cal F}(-\frac{V'}{V})\equiv F(V,V').$ Let
$h=V'$, then
\begin{equation}\frac{dh}{dV}=\frac{1}{h}F(V,h)=\frac{h}{V}{\cal F}(-\frac{h}{V})\end{equation}
Now Eq.(11) is a one-order differential equation of $h$ and $V$.
Figuring out $h(V)$, the potential can be solved from equation
$V'(\phi)=h(V(\phi))$. \par Here we introduce another easier
approach to get the potential $V(\phi)$. We start with
$\frac{d\lambda}{dV}=\frac{d\lambda}{d\phi}\frac{d\phi}{dV}=-\frac{d(V'/V)}{d\phi}\frac{1}{V'}=-\frac{1}{V'}\frac{V''V-V'^2}{V^2}$.
Using the definition of $\lambda$ and $\Gamma$, and the Eq.(8), we
get a one-order differential equation of $\lambda$ and $V$:
\begin{equation}\frac{d\lambda}{dV}=\frac{\lambda}{V}f(\lambda)\end{equation}
Integrating out $\lambda=\lambda(V)$, using the definition of
$\lambda$, then we have following differential equation of
potential:
\begin{equation}\frac{dV}{V\lambda(V)}=-d \phi\end{equation}
So Eq.(12) and Eq.(13) give the route to obtain the potential
$V=V(\phi)$. As far as we know, there are too many investigations to
the two-dimension autonomous system (where the potential is
exponential) but have not general investigations to the dynamical
properties of three dimensional autonomous system. It is maybe very
interesting to consider this issue. We know that previous exactly
analysis to the critical points of the quintessence model are based
on a concrete form of potential (i.e., the exponential form). In
this case it is not easy to distinguish which critical points are
common to all the quintessence models and which are only related to
the special potentials. In view of what we mention above, we will
take a new route in next section to investigate the critical points
of the autonomous system with an arbitrary function of $f(\lambda)$.
Furthermore, the results can be easily applied to any other concrete
potentials as long as they can be solved from Eqs.(12-13).
\section{Critical Points and their Properties}
\par It is easily seen from Eq.(9) that $\lambda=0, x=0$ or
$f(\lambda)=0$ can make $d\lambda/dN=0$ respectively. The critical
points listed in TABLE 1 can be found from the Eqs.(5,6,9) after
setting $dx/dN=dy/dN=d\lambda/dN=0$. The properties of each critical
point are determined by the eigenvalues of the Jacobi matrix of the
three-dimension autonomous system. For a general three-dimension
autonomous system:
\begin{equation}\left\{
\begin{array}{ll}\frac{dx}{dN}=f_1(x, y, \lambda)\\
\frac{dy}{dN}=f_2(x, y, \lambda)\\
\frac{d\lambda}{dN}=f_3(x, y, \lambda)\end{array}\right.
\end{equation}
\par The function $f_1, f_2$ and $f_3$ are only the function of $x, y,
\lambda$, no variable $N$ and other variables, we call this
dynamical system as autonomous system. If $f_1, f_2$ and $f_3$ are
only a linear combination of $x, y, \lambda$,  Eq.(14) is linear
autonomous system. Its critical points ($x_c, y_c, \lambda_c$) can
be found from the set of functions $f_1= f_2=f_3=0$. Obviously,
Eqs.(5,6,9) is not a linear autonomous system. However, the local
behavior of the nonlinear autonomous system near a critical point
can be deduced by linearizing the nonlinear system about this point
and be studied using the linear autonomous system analysis method.
The properties of each critical point are determined by the
eigenvalues of the Jacobi matrix ${\cal A}$, where
\begin{equation}{\cal A}=\left[ \begin{array}{lll}
\partial f_1(x, y,\lambda)/\partial x & \partial f_1(x,x y, \lambda)/\partial y &\partial f_1(x, y, \lambda)/\partial \lambda \\
\partial f_2(x, y,\lambda)/\partial x & \partial f_2(x, y, \lambda)/\partial y &\partial f_2(x, y, \lambda)/\partial \lambda \\
\partial f_3(x, y,\lambda)/\partial x & \partial f_3(x, y, \lambda)/\partial y &\partial f_3(x, y, \lambda)/\partial \lambda
\end{array}\right]_{(x_c, y_c, \lambda_c)}\end{equation}
\footnotetext{$^1$ Actually "critical point" in this paper is also
called the "equilibrium point" in mathematics or "fixed point" in
some physical literatures. a hyperbolic critical(equilibrium) point
is the critical(equilibrium) point which has no eigenvalues with
zero real part.} \footnotetext{$^2$ i.e., its eigenvalues exist zero
value or have zero real parts.} For a hyperbolic critical
point$^{1}$, if all the eigenvalues of ${\cal A}$ or the real part
of these eigenvalues are negative, the critical point is stable.
This is to say, as long as one of the eigenvalues or the real part
of these eigenvalues is positive, the critical point must be
unstable. However, if the critical point of nonlinear autonomous
system is a nonhyperbolic point$^{2}$ and the rest of its
eigenvalues having negative real part, the properties of this point
can not be simply determined by linearization method and need to
resort to other more complicated methods[50]. From TABLE 1, we can
see that point $P_4$ is just this kind of point. In previous
literatures[30, 31, 37, 51], the authors generally neglected this
nonhyperbolic point when they met it. In fact this point also has
the important cosmological implication as other critical points and
should not be ignored. We will explore the properties of this
nonhyperbolic point $P_4$ in our paper using the center manifold
theorem[50] (The full analysis process is given in the Appendix). We
list all the points and their properties in the following TABLE 1.
Note that we have neglected the cases with $y<0$ since the system is
symmetric under the reflection $(\lambda, x, y)\rightarrow(\lambda,
x, -y)$ and time reversal $t\rightarrow -t$.

{ \begin{center}\begin{tabular}{|c|c|c|c|}\hline &$(\lambda_c, x_c,
y_c)$& eigenvalues & Stability \\ \hline $P_1$&$(0, 1, 0)$&
$3(1-w_b), 3, 0$& unstable node \\ \hline $P_2$&$(0,-1, 0)$&
$3(1-w_b), 3, 0$& unstable node\\ \hline $P_3$&$(0, 0, 0)$&
$-3(1-w_b)/2, 3\gamma/2, 0$& saddle point
\\ \hline
$P_4 $ &$(0, 0, 1)$ & $-3, -3\gamma, 0$ & stable node for \\
\space &\space &\space &$f(0)>0$ \\ \hline
 $P_5$&$(\lambda_{a}, 0, 0)$&$-3(1-w_b)/2, 3\gamma/2, 0$ &
saddle point\\ \hline $P_6$&$(\lambda_{*}, 0, 0)$&
$-3(1-w_b)/2,3\gamma/2, 0$& saddle point\\ \hline
$P_7$&$(\lambda_{*}, 1, 0)$& $-\sqrt 6 \lambda_{*}^2df_{*},3(1-w_b),
\frac{1}{2}(6-\sqrt 6 \lambda_{*})$& saddle point \\ \hline
$P_8$&$(\lambda_{*}, -1, 0)$& $\sqrt 6 \lambda_{*}^2df_{*},3(1-w_b),
\frac{1}{2}(\sqrt 6 \lambda_{*}+6)$& saddle point \\ \hline
$P_9$&$(\lambda_{*}, \frac{\sqrt6}{6}\lambda_{*},
\sqrt{1-\frac{1}{6}\lambda_{*}^2})$& $\frac{1}{2}(\lambda_{*}^2-6),
\lambda_{*}^2-3\gamma, -\lambda_{*}^2 \lambda_{*}df_{*} $& Eq.(16)
\\ \hline $P_{10}$&$(\lambda_{*}, \frac{\sqrt6\gamma}{2
\lambda_{*}},\frac{\sqrt{6\gamma(1-w_b)}}{2\lambda_{*}})$&$
-3\lambda_{*}\gamma df_{*}, \frac{3}{4}(w_b-1)$&
Eq.(17) \\& &$\pm\frac{3\sqrt{(1-w_b)}}{4\lambda_{*}}\sqrt{24\gamma^2-(9\gamma-2)\lambda_{*}^2}$&\\
\hline\end{tabular}\end{center}}
\par Where $f(0)$ is the value of function $f(\lambda)$ at $\lambda=0$, $df_{*} \equiv
\frac{df(\lambda)}{d\lambda}|_{\lambda_{*}}$. We limit the range of
$w_b(=\gamma-1)$ as $0\leq w_b<1$, $w_b=0$ for matter and $1/3$ for
radiation. $\lambda_a$ means an arbitrary value and $\lambda_{*}$ is
the value which makes $f(\lambda_{*})=0$. { So points $P_{7-10}$
appear only if the function $f(\lambda)$ can be zero for one or more
values of $\lambda_{*}$. Here we simply consider that only one value
$\lambda_{*}$ makes the function $f(\lambda)$ zero.
\par However, readers should keep in mind that, to make $d\lambda/dN=0$ in Eq.(9), we let $\lambda=0, x=0$
and $f(\lambda)=0$ separately, and then find out all the points
listed in TABLE 1. But we do not consider one special case that
$\lambda^2f(\lambda)\neq 0$ and then $d\lambda/dN \neq 0$ when
$\lambda=0$. In this case, $P_1$ and $P_2$ are no more critical
points. For example, the product $\lambda^2 f(\lambda)= \frac{V_0
\alpha^2}{\Lambda}\neq 0$ even if $\lambda=0$ for the potential
$V(\phi)=V_0[\cosh(\alpha\phi)-1]+\Lambda$. So the necessary
condition for the existence of equilibrium points with $x \neq 0$ is
$\lambda^2f(\lambda)=0$.
\par $\lambda_{*}^2<6$ is the condition for critical point $P_9$ to exist and Eq.(16) is the
condition for $P_9$ to be a stable node.}
\begin{equation}\lambda_{*}^2<3\gamma  \ and \  \lambda_{*} df_{*}>0\end{equation}
\par $\lambda_{*}^2>3\gamma$ is the condition for critical point $P_{10}$ to exist and Eq.(17) is its stable condition.
\begin{equation}\begin{array}{lll}24 \gamma^2/(9\gamma-2)>\lambda_{*}^2>3\gamma \ and\  \lambda_{*} df_{*}>0\ \ for\ P_{10}\ being\ a\ stable\ node
\\ \lambda_{*}^2>24 \gamma^2/(9\gamma-2) \ and\  \lambda_{*} df_{*}>0\ \ for\ P_{10}\ being\ a\ stable\ spiral\end{array}\end{equation}
\par the density parameter of $\phi$ field and its equation of state are,
respectively:
\begin{equation}\Omega_{\phi}=x^2+y^2\end{equation}
\begin{equation}w_{\phi}=\frac{x^2-y^2}{x^2+y^2}\end{equation}
 In order to investigate the expansive behavior of scale factor $a$, we also represent the
decelerating factor:
\begin{equation}\begin{array}{lll}
q &=&-\frac{\ddot{a}a}{\dot{a}^2}=-\frac{\ddot
a}{a}/H^2=\frac{\sum{(1+3w_i)\rho_i}}{2\sum{\rho_i}}=\frac{1}{2}\sum(1+3w_i)\Omega_i
\\\\&=&\frac{3}{2}[(1-w_b)x^2-(1+w_b)y^2+(w_b+\frac{1}{3})]\end{array}\end{equation}
We list the other properties of these critical points in TABLE 2. {{
\begin{center}\begin{tabular}{|c|c|c|c|c|c|} \hline &$(\lambda_c,
x_c, y_c)$&$w_{\phi}$& $\Omega_{\phi}$ & decelerating factor(q)\\
\hline
$P_1$&$(0, 1, 0)$&1& 1& 2  \\
\hline
$P_2$&$(0, -1, 0)$&1& 1& 2  \\
\hline
$P_3$&$(0, 0, 0)$& Undefined & $0$& $(3w_b+1)/2$ \\
\hline
$P_4$&$(0, 0, 1)$&$-1$& 1& $-1$   \\
\hline
$P_5$&$(\lambda_{a}, 0, 0)$& Undefined & $0$& $(3w_b+1)/2$   \\
\hline
$P_6$&$(\lambda_{*}, 0, 0)$& Undefined & $0$& $(3w_b+1)/2$   \\
\hline
$P_7$&$(\lambda_{*}, 1, 0)$&1& 1& 2 \\
\hline
$P_8$&$(\lambda_{*}, -1, 0)$&$1$& $1$& 2   \\
\hline
$P_9$&$(\lambda_{*}, \frac{\sqrt6}{6}\lambda_{*}, \sqrt{1-\frac{1}{6}\lambda_{*}^2})$&$\lambda_{*}^2/3-1$&$1$& $\lambda_{*}^2/2-1$  \\
\hline
$P_{10}$&$(\lambda_{*}, \frac{\sqrt6\gamma}{2
\lambda_{*}},\frac{\sqrt{6\gamma(1-w_b)}}{2\lambda_{*}})$&$w_b$ & $3\gamma/\lambda_{*}^2$& $(3w_b+1)/2$ \\
\hline
\end{tabular}\end{center}}

\section{Cosmological Implications}
\hspace*{16 pt} After giving all the critical points and their
properties of the three-dimension autonomous system, we will
investigate their cosmological implications. We will show some
interesting results which have not been found previously in other
literatures. Moreover, we will also response to the questions we
have proposed in Section 1. Investigating three-dimension autonomous
system instead of the two-dimension autonomous system can help us
consider more potentials which can not be investigated via
two-dimension autonomous system. Moreover, from the view of
three-dimension system, we can gain a more deeply understanding than
from the two-dimension system. For example, we will point out which
critical points are the critical points for all quintessence and
which are only relative to the concrete potentials. We can find from
TABLE 1 and TABLE 2 that: {{ Though the stability of Points $P_{1,
2}$ does not depend on the form of concrete potentials, Points
$P_{1, 2}$ only exist when $\lambda^2f(\lambda)= 0$ at $\lambda =0$
. Points $P_{3, 5, 6}$ always exist for all quintessence models and
their stability are regardless of the form of concrete potentials.
Point $P_4$ is also the critical point for all quintessence, but its
stability depends on the form of concrete potentials. Points
$P_{7-10}$ and their properties are closely connected to the
concrete potentials since the value of $\lambda_{*}$ is determined
by the form of $f(\lambda)$.  points $P_{7-10}$ are even inexistence
if $f(\lambda) \neq0$ for any $\lambda$.
\par Of all the points, only Points $P_{3,5,6}$ are independent of the function $f(\lambda)$.} In fact, they
 have the same properties and can be considered as one point.
They are {{ saddle points} which tell us that the barotropic fluid
dominated solution $(\lambda_c=0, x_c=0, y_c=0)$ where
$\Omega_{\phi}=0$ is unstable. {{However, even though they are
unstable, the phase space trajectories may evolve in the vicinity of
the barotropic fluid dominated solution for a quite long time and
then leaves this state to approach to the possible future
attractor.} However, if $\gamma=0$, these points are found to be a
stable attractor and can be used to alleviate the relic density
problem in inflation model[37].
\par Four of the critical points ($P_{1, 2}$$(\lambda_c=0, x_c=\pm 1,
y_c=0)$ and $P_{7, 8}$($\lambda_c=\lambda_{*}, x_c=\pm 1, y_c=0)$)
are all unstable   nodes, which correspond to the solutions where
the universe is dominated by the kinetic energy of the scalar field
($\Omega_{\phi}=1$) with a stiff equation of state ($w_{\phi}=1$).
\par In fact, we can conclude above results with one brief sentence (see TABLE 1): {\it  all the critical points with $y_c$
being zero are not stable points.} It tells us that, under the
potential we considered here, the cosmological solution with the
potential energy eventually evolving to zero will never be the final
state of our universe.} This is a quite interesting result since we
know that the universe will never undergo a regime of accelerating
expansion if there is no potential energy in quintessence models.
\par Therefore, there are only three critical points $P_{4, 9, 10}$  which
correspond to possible late-time attractor solutions. We will study
their properties and cosmological implications in more detail.

\par Points $P_{4, 9}$ are both scalar field dominated solutions with
$\Omega_{\phi}=1$.  Comparing with point $P_4$, P$_9$ is the
well-known scalar field dominated solution which exists for
$\lambda_{*}^2<6$. TABLE 1 has shown that this scalar field
dominated solution is a later-time attractor in the presence of a
barotropic fluid if we have $\lambda_{*}^2<3\gamma \ and \
\lambda_{*} df_{*}>0$. This solution will give an accelerating
universe if $\lambda_{*}^2<2  \ and \  \lambda_{*} df_{*}>0$. For
example, $f(\lambda)=\frac{1}{n}-\frac{n\sigma^2}{\lambda^2}$
corresponds to $V(\phi)=\frac{V_0}{[cosh(\sigma\phi)]^n}$. Obviously
we have $\lambda_{*}=\pm|n\sigma|$ and
$df_{*}=\frac{2n\sigma^2}{\lambda_{*}^3}$. The scalar field
dominated solution with potential
$V(\phi)=\frac{V_0}{[cosh(\sigma\phi)]^n}$ is a late-time attractor
if $n^2\sigma^2<3 \gamma \ and \ \frac{2n
\sigma^2}{\lambda_{*}^2}>0$. In addition, this solution admits an
accelerating expansion of universe if $n^2\sigma^2<2 \ and \
\frac{2n \sigma^2}{\lambda_{*}^2}>0$. Noted that the point $P_9$
means two stable critical points ($\lambda_c=\pm|n\sigma|,
x_c=\frac{\pm\sqrt6}{6}|n\sigma|,
y_c=\sqrt{1\mp\frac{1}{6}n^2\sigma^2}$) in this case .

\par $P_{10}$ is the scaling solution where neither the
scalar field nor the barotropic fluid entirely dominates the
universe. $P_{10}$ is a stable node for $24
\gamma^2/(9\gamma-2)>\lambda_{*}^2>3\gamma \ and\ \lambda_{*}
df_{*}>0$ and a stable spiral for $\lambda_{*}^2>24
\gamma^2/(9\gamma-2) \ and\  \lambda_{*} df_{*}>0$. So $P_9$ and
$P_{10}$ can not be stable simultaneously. The scaling solution has
drawn a lot of attentions since it can alleviate the coincidence
problem of dark energy. Many potentials have been proposed to give a
scaling evolution regime[35, 36, 52-65]. Here we give a sufficient
condition for a potential to possess a scaling solution, that is,
{\it as long as $f(\lambda)$ equals zero for one or more values of
$\lambda(=\lambda_{*})$ and these $\lambda_{*}$ also satisfy
Eq.(17), then there must exist a scaling solution with
$\Omega_{\phi}=3\gamma/\lambda_{*}^2$}. Obviously many potentials
which satisfy this condition exist, such as the potential
$V(\phi)=\frac{V_0}{[cosh(\sigma\phi)]^n}$ which corresponds to
$f(\lambda)=\frac{1}{n}-\frac{n\sigma^2}{\lambda^2}$, the potential
$V(\phi)=\frac{V_0}{(\eta+e^{-\alpha\phi})^{\beta}}$ which
corresponds to $f(\lambda)=\frac{1}{\beta}+\frac{\alpha}{\lambda}$
and so on. Our condition includes the potentials in Ref[49] where
the authors found that every positive and monotonous potential which
was asymptotically exponential yielded a scaling solution.  Our
result is also not contradiction to the statement in literature[66,
67] where they assumed a scaling solution like $P_{10}$ and found
the potential was unique the exponential form. This exponential
potential is explicitly figured out from the assumption and the
evolution of universe with this potential is always the scaling
solution(see Eq.(18) in literature[66]) while $P_{10}$ being a
stable point means that all the evolution of the universe with a
class of potentials which satisfy Eq.(17) will all approach the
scaling solution finally. It is just an asymptotic behavior at late
time. Unfortunately, for the scaling solution of $P_{10}$, the state
equation of dark energy $w_{\phi}$ equals $w_m$ and therefore there
does not exist the accelerating expansion if $w_m$ is larger than
zero. However, authors had obtained the exact quintessence potential
$V(\phi)=\frac{1-w_{\phi}}{2}\rho_{\phi_0}[\sqrt{\frac{\Omega_{m0}}{\Omega_{\phi
0}}}sinh(\frac{3(w_m-w_{\phi})}{2\sqrt{3(1+w_{\phi})}}\frac{\phi-\phi_{in}}{m_{pl}})]^{-2(1+w_{\phi})/(w_m-w_{\phi})}
$, which admited a scaling solution with $w_{\phi}\neq w_m \ and \
\Omega_{\phi} \neq 0$[66]. With this potential, in principle, we can
obtain a scaling solution with an accelerating expansion of the
universe.

\par Finally, let us consider the point $P_4$, which is a
de-Sitter-like dominant attractor with $\Omega_{\phi}=1 \ and \
w_{\phi}=-1$. The condition for $P_4$ being a stable point is that
the value of $f(\lambda)$ when $\lambda=0$ must be larger than
zero(i.e., $f(0)>0$, see appendix for details). So generally
speaking, $P_4$ and $P_{9}$(or $P_{10}$) also can not be stable
simultaneously. However, there may exist the possibility for some
potentials that their values at $\lambda=0$ is
 larger than zero but equals zero for some others $\lambda_{*}$ ($\lambda_{*}\not=0$), then this region of $\lambda$ in the phase space of the
three dynamical autonomous system will lie in the basin of the
attractor $P_{10}$. That means, in this case, there can exist two
stable critical points simultaneously, but this is not to say that
the universe can evolve continuously  from one stable critical point
to another one. Based on this fact, the author proposed a scenario
of universe which could evolve from a scaling attractor to a
de-Sitter-like attractor by introducing a field whose value changed
a certain amount in a short time[47]. In fact, we can also obtain
these two asymptotical evolutions if the potential $V(\phi)$ can be
approximated to two different potentials when $\phi$ evolves to
different range, one admits the scaling solution and another one
admits the de-Sitter-like solution[53, 58, 61]. For these
potentials, the exit of the cosmological evolution from one
attractor solution to another attractor is quite natural, but the
explanation of why we have these special potentials is not quite
natural.
\section{Conclusion}
In this paper, we extend the autonomous dynamical system analysis of
the canonical scalar field from 2-D to three-dimension by
considering the potential parameter $\Gamma$ as a function of
another potential parameter $\lambda$. There are ten critical points
in all: { three of these points ($P_{3, 5, 6}$) are general points
which are possessed by all quintessence models regardless of the
form of potentials and the rest points, with their existence or/and
stability,   are closely connected to the concrete potentials.} We
surprisingly find that, apart from the exponential potential, there
are a large number of potentials which can give the scaling solution
when the function $f(\lambda)(=\Gamma(\lambda)-1)$ equals zero for
one or some values of $\lambda$ and the parameter $\lambda$
satisfies the condition Eq.(16) or Eq.(17) at the same time. We give
the explicit expression to derive these potentials $V(\phi)$ from
$f(\lambda)$. We find that, if some conditions are satisfied, the
de-Sitter-like dominant point $P_4$ and the scaling point $P_9$( or
$P_{10}$) can simultaneously be stable, but $P_9$ and $P_{10}$ can
not be stable at one time. As we have seen, the autonomous dynamical
systems analysis is a very powerful tool which helps us extract
useful cosmological information without solving the complicated
background equations. Our method extends the analysis from
two-dimensional autonomous dynamical system to three-dimension,
which makes us be able to research a large number of potentials
beyond the exponential potential. This method is quite effective and
may be applied to a broad class of dark energy models studied in
literature[46], including coupled quintessence, (coupled-)phantom
scalar field, k-essence and even generalized background $H^2
\propto\rho_T^n$.
\par {However, we should point out that our approach also has its drawbacks. First, as we have mentioned above: the approach
can not be applied for the potentials for which the function
$\Gamma=V V''/(V')^2$ can not be written as an explicit function of
the variable $\lambda$. Second, the variable $\lambda$ is undefined
if the potentials vanish at its minimum, so the approach can not be
applied for the potentials which vanish at its minimum. But, in
despite of the second problem, it is actually not a fatal drawback.
On one hand, the minimum of a potential is always associate with the
late-time cosmological dynamics(future attractors). It is quite easy
to discuss this special equilibrium point separately if we know a
given potential has minimum(It is usually not difficult to find out
the minimum of a given function). On the other hand, we can still
use our approach to analyze all the critical points $P_{1-10}$ since
$\lambda$ is well-defined around these critical points. We take the
potential $V(\phi)=V_0[cosh(\alpha \phi) -1]$ for example, this
potential has a minimum value $0$ at $\phi=0$($\lambda$ has no
definition at $\phi=0$). We can investigate the critical point
corresponding to this minimum separately. The explicit function
about this potential is
$f(\lambda)=\frac{1}{2}(\frac{\alpha^2}{\lambda^2}-1)$. Obviously,
the point corresponding to the potential's minimum does not appear
in the Table 1. We can still discuss the properties of the critical
points $P_{1-10}$  even if the variable $\lambda$ has no definition
sometime.}
\section{Acknowledgement}
\par W.Fang owns the great improvement of our paper to the anonymous referees.} W.Fang would like to thank Shuang-Yong Zhou, Dr.Hong Tu, Dr.Zhu Chen and Prof.Yun-Gui
Gong for useful discussion. W.Fang also kindly thanks the Abdus
Salam International Centre for Theoretical Physics(ICTP) for the
support to attend the summer school in cosmology when part of the
work has been finished. This work is partly supported by Shanghai
Municipal Science and Technology Commission under Grant
No.07dz22020, Shanghai Normal University under Grant No.DZL712,
No.DKL934, No.PL905 and Natural Science Foundation of Jiangsu
Province under Grant No.07KJD140011.
\\
\\
{\Large \bf Appendix} \\
\par {\small In section 3, we pointed out that if the eigenvalues of
Jacobi matrix had one or more eigenvalues with zero real parts while
the rest of the eigenvalues had negative real parts, then
linearization fails to determine the stability properties of this
critical point. From TABLE 1 we realize that point $P_4$ is just
such point, so in this Appendix we will show you that how we get the
stable condition of $P_4$ from the center manifold theorem. The
point $P_4$ is $ (\lambda_c=0, x_c=0, y_c=1)$} and its three
eigenvalues are
 $(0, -3, -3(1+w_m))$. Firstly, we transfer $P_4$ to $P_4'$ $(\lambda_c=0, x_c=0,
 Y_c=y_c-1=0$) for convenience. In this case,  Eqs.(5-7) can be rewritten as:
\begin{equation}\frac{d\lambda}{dN}=-\sqrt{6} \lambda^2f(\lambda)x \end{equation}
\begin{equation}\frac{dx}{dN}=-3x+\frac{1}{2}\sqrt{6}\lambda+\frac{1}{2}\sqrt{6}\lambda Y^2+\sqrt{6}\lambda Y+\frac{3}{2}x^3(1-w_m) -\frac{3}{2}(1+w_m)xY^2-3(1+w_m)xY\end{equation}
\begin{equation}\frac{dY}{dN}=-3(1+w_m)Y-\frac{1}{2}\sqrt{6}\lambda x(Y+1)+\frac{3}{2}(1-w_m)x^2Y-\frac{3}{2}Y^3
-\frac{3}{2}(3+w_m) Y^2+\frac{3}{2}(1-w_m)x^2\end{equation} \par
Noted that $\{\lambda, x, Y\}$ in Eqs.(21-23) are very small
variables around point $(\lambda_c=0, x_c=0, Y_c=0$. So the function
$f(\lambda)$ in Eq.(21) should be taken the Taylor series in
$\lambda$: $f(\lambda)=f(0)+f^{1}(0)
\lambda+\frac{f^{2}(0)}{2!}\lambda^2+...$, where  $f^{n}(0)$ is the
value of $\frac{d^n f(\lambda)}{d \lambda^n}$ when $\lambda=0$.
\par We can write down the Jacobi matrix ${\cal A}$ of dynamical
system Eqs.(21-23):\begin{equation} {\cal A}= \left [
\begin{array}{lll}
\ \ 0 & \ 0 & \ \ \ \ \ \ \ \ 0 \\ \frac{1}{2}\sqrt{6} & -3 &\ \ \ \
\ \ \ \ 0 \\ \ \ 0 & \ 0 &-3(1+w_m)\end{array} \right
]\end{equation}
\par The eigenvalues of ${\cal A}$ and the corresponding eigenvectors
are:\begin{equation}\{0,\ \ [1, \ \ \frac{\sqrt{6}}{6}, 0]\};\ \ \ \
\ \{-3,\ \  [0, 1, 0]\}; \ \ \ \ \ \{-3(1+3w_m),\ \ [0, 0,
1]\}\end{equation} Let ${\cal M}$ be a matrix whose columns are the
eigenvectors of ${\cal A}$, then we can write down ${\cal M}$ and
its inverse matrix ${\cal T}$:
\begin{equation} {\cal M}= \left [ \begin{array}{lll}
\ 1 & 0 & 0 \\ \frac{\sqrt{6}}{6} & 1 & 0 \\  \ 0 &  0 & 1
\end{array} \right ], \ \ \ \ \ \ \ {\cal T}={\cal M}^{-1}= \left [
\begin{array}{lll}
\ \ 1 & 0 & 0 \\ -\frac{\sqrt{6}}{6} & 1 & 0 \\  \ \ 0 &  0 & 1
\end{array} \right ]\end{equation}
 Using the similarity transformation ${\cal T}$ we can transform ${\cal
 A} $ into a block diagonal matrix, that is,
\begin{equation} {\cal T}{\cal A}{\cal T}^{-1}=\left [ \begin{array}{lll}
0 & \ 0 & \ \ \ \ \ \ \ \ 0 \\ 0 & -3 &\ \ \ \ \ \ \ \ 0 \\ 0 & \ 0
&-3(1+w_m)\end{array} \right ]=\left [ \begin{array}{ll}{\cal A}_1 &
\ 0 \\ 0 &  {\cal A}_2 \end{array} \right ]\end{equation} where all
eigenvalues of ${\cal A}_1$ have zero real parts and all eigenvalues
of ${\cal A}_2$ have negative real parts. We put a change of
variables:\begin{equation} \left [\begin{array}{l}\lambda' \\ x' \\
Y'\end{array} \right ] ={\cal T} \left [\begin{array}{l}\lambda \\ x
\\ Y\end{array}\right ] =\left [\begin{array}{l}\ \ \ \ \ \ \lambda \\-\frac{\sqrt{6}}{6}\lambda+x \\
\ \ \ \ \ \ Y\end{array} \right ]\end{equation} Then we can rewrite
the dynamical system Eqs.(21-23) in the form of new variables:
\begin{equation}\frac{d\lambda'}{dN}=\frac{d\lambda}{dN}=f_1(\lambda', x', Y')\end{equation}
\begin{equation}\frac{dx'}{dN}=-\frac{\sqrt{6}}{6} \frac{d\lambda}{dN}+\frac{dx}{dN}=f_2(\lambda', x', Y')\end{equation}
\begin{equation}\frac{dY'}{dN}=\frac{dY}{dN}=f_3(\lambda', x', Y')\end{equation}
the detail forms of $f_1(\lambda', x', Y'), f_2(\lambda', x', Y'),
f_3(\lambda', x', Y')$ are easily obtained after we substitute the
transformation $\lambda=\lambda'$, $x=\frac{\sqrt{6}}{6}\lambda'+x'$
and $Y=Y'$ into the right hand of Eqs.(21-23). According to the
center manifold theorem, the stable condition of dynamical system
Eq.(21-23), i.e., the stability of $P_4$ will be finally determined
by the following simple reduced system:
\begin{equation}\frac{d\lambda'}{dN}=\frac{d\lambda}{dN}=-\lambda'^3f(0)=-\lambda^3f(0)\end{equation}
$f(0)$ is the value of function $f(\lambda)$ at $\lambda=0$. This
simple one-dimensional dynamical system Eq.(32) is stable if
$f(0)>0$.
\par So we conclude that $P_4$ is a stable de-Sitter-like dominant
attractor when $f(0)>0$, just as shown in TABLE 2.
\\
\\
{\noindent\Large \bf References} \small{\begin{description}
\item {[1]}{B.Ratra and P.J.E.Peebles, Phys.Rev.D\textbf{37}, 3406(1988).}
\item {[2]}{P.J.E.Peeble, B.Ratra, Astrophys.J \textbf{325}, L17(1988).}
\item {[3]}{R.Caldwell et al., Phys.Rev.Lett \textbf{80}, 1682(1998).}
\item {[4]}{J.S.Bagla, H.K.Jassal and T.Padmamabhan, Phys.Rev.D\textbf{67}, 063504(2003).}
\item {[5]}{L.Amendola et al., Phys.Rev.D\textbf{74}, 023525(2006).}
\item {[6]}{S.Nojiri, S.D.Odintsov and M.Sasaki, Phys.Rev.D\textbf{70}£¬043539(2004).}
\item {[7]}{C.Wetterich£¬Nucl.Phys.B\textbf{302}, 668(1998).}
\item {[8]}{I.Zlatev, L.Wang and P.J.Steinhardt£¬Phys.Rev.Lett\textbf{82}, 896(1999).}
\item {[9]}{A.Sen, JHEP \textbf{0204}, 048(2002).}
\item {[10]}{C.Armendariz-Picon et al., Phys.Lett.B\textbf{458}, 209(1999).}
\item {[11]}{X.Z.Li, J.G.Hao and D.J.Liu, Class.Quantum Grav.\textbf{19}, 6049(2002).}
\item {[12]}{A.Feinstein, Phys.Rev.D\textbf{66}, 063511(2002).}
\item {[13]}{A.Frolov, L.Kofman and A.Starobinsky, Phys.Lett.B\textbf{545}, 8(2002).}
\item {[14]}{C.Armendariz-Picon et al., Phys.Rev.Lett\textbf{85}, 4438(2000)}
\item {[15]}{T.Chiba, Phys.Rev.D\textbf{66}, 063514(2002).}
\item {[16]}{L.P.Chimento, Phys.Rev.D\textbf{69}, 123517(2004).}
\item {[17]}{A.Melchiorri et al., Phys.Rev.D\textbf{68}, 043509(2003).}
\item {[18]}{R.R.Caldwell, Phys.Lett.B\textbf{545}, 23(2002).}
\item {[19]}{T.Chiba, T.Okabe and M.Yamaguchi, Phys.Rev.D\textbf{62}, 023511(2000).}
\item {[20]}{L.Amendola, Phys.Rev.Lett\textbf{93}, 181102(2004)}
\item {[21]}{S.M.Carroll, M.Hoddman and M.Trodden, Phys.Rev.D\textbf{68}, 023509(2003).}
\item {[22]}{X.Z.Li and J.G.Hao, Phys.Rev.D\textbf{69}, 107303(2004).}
\item {[23]}{L.R.Abramo, F.Finelli and T.S.Pereira, Phys.Rev.D\textbf{70}, 063517(2004).}
\item {[24]}{H.Q.Lu, Int.J.Mod.Phys.D\textbf{14}, 355(2005).}
\item {[25]}{M.R.Garousi, M.Sami and S.Tsujikawa, Phys.Rev.D\textbf{71}, 083005(2005).}
\item {[26]}{M.Novello, M.Makler, L.S.Werneck and C.A.Romero, Phys.Rev.D\textbf{71}, 043515(2005).}
\item {[27]}{W.Fang, H.Q.Lu and Z.G.Huang, Class.Quantum Grav.\textbf{24}, 3799(2007).}
\item {[28]}{W.Fang, H.Q.Lu, B.Li and K.F.Zhang, Int.J.Mod.Phys.D\textbf{15}, 1947(2006).}
\item {[29]}{W.Fang, H.Q.Lu, Z.G.Huang and K.F.Zhang, Int.J.Mod.Phys.D\textbf{15}, 199(2006).}
\item {[30]}{L.Amendola, Phys.Rev.D\textbf{60}, 043501(1999).}
\item {[31]}{L.Amendola, Phys.Rev.D\textbf{62}, 043511(2000).}
\item {[32]}{W.Hao, R.G.Cai and D.F.Zeng, Class.Quant.Grav\textbf{22}, 3189(2005). }
\item {[33]}{Z.K.Guo, Y.S.Piao, X.M.Zhang and Y.Z.Zhang, Phys.Lett.B\textbf{608},177(2005).}
\item {[34]}{B.Feng£¬X.L.Wang and X.M.Zhang£¬Phys.Lett. B\textbf{607}£¬35-41(2005).}
\item {[35]}{A.A.Coley and R.J.van den Hoogen, Phys.Rev.D\textbf{62}, 023517(2000).}
\item {[36]}{S.A.Kim and A.R.Liddle and S.Tsujikawa, Phys.Rev.D\textbf{72}, 043506(2005).}
\item {[37]}{E.J.Copeland, A.R.Liddle and D.Wands, Phys.Rev.D\textbf{57}, 4686(1998).}
\item {[38]}{J.G.Hao and X.Z.Li, Phys.Rev.D\textbf{67}, 107303(2003).}
\item {[39]}{J.G.Hao and X.Z.Li, Phys.Rev.D\textbf{70}, 043529(2004).}
\item {[40]}{Z.K.Guo, R.G.Cai and Y.Z.Zhang , JCAP\textbf{0505}, 002(2005).}
\item {[41]}{W.Fang, H.Q.Lu and Z.G.Huang, Int.J.Theor.Phys\textbf{46}, 2366(2007).}
\item {[42]}{Z.K.Guo, Y.S.Piao and Y.Z.Zhang, Phys.Lett.B\textbf{568}, 1-7(2003).}
\item {[43]}{Z.K.Guo, Y.S.Piao£¬R.G.Cai and Y.Z.Zhang, Phys.Lett.B\textbf{576}, 12-17(2003).}
\item {[44]}{A.de la Macorra and G.Piccinelli, Phys.Rev.D\textbf{61}, 123503(2000).}
\item {[45]}{S.C.C.Ng, N.J.Nunes and F.Rosati, Phys.Rev.D\textbf{64}, 083510(2001).}
\item {[46]}{E.J.Copeland, M.Sami and S.Tsujikawa, Int.J.Mod.Phys.D\textbf{15}, 1753(2006).}
\item {[47]}{S.Y.Zhou, Phys.Lett.B\textbf{660}, 7-12(2008).}
\item {[48]}{R.R.Caldwell, R.Dave and P.J.Steinhardt, Phys.Rev.Lett\textbf{80}, 1582(1998).}
\item {[49]}{Ana Nunes and Jose P.Mimoso, gr-qc/0008003.}
\item {[50]}{Hassan K.Khalil, Nonlinear Systems(Second Edition),Prentice Hall(1996), p167-p177.}
\item {[51]}{B.Gumjudpai, T.Naskar, M.Sami and S.Tsujikawa, JCAP\textbf{506}, 007(2005)}
\item {[52]}{S.Mizuno, S.J.Lee and E.J.Copeland, Phys.Rev.D\textbf{70},043525(2004).}
\item {[53]}{T.Barreiro, E.J.Copeland and N.J.Nunes, Phys.Rev.D\textbf{61}, 127301(2000).}
\item {[54]}{A.A. Sen and S.Sethi, Phys.Lett.B\textbf{532}, 159(2002).}
\item {[55]}{I.P.Neupane, Class.Quant.Grav.\textbf{21}, 4383(2004).}
\item {[56]}{I.P.Neupane, Mod.Phys.Lett.A\textbf{19}, 1093(2004).}
\item {[57]}{L.Jarv, T.Mohaupt and F.Saueressig, JCAP\textbf{0408}, 016(2004).}
\item {[58]}{V.Sahni and L.M.Wang, Phys.Rev.D\textbf{62}, 103517(2000)..}
\item {[59]}{T.Matos and L.A.Urena-Lopez, Class.Quant.Grav.\textbf{17}, L75(2000).}
\item {[60]}{W.Hu, R.Barkana and A.Gruzinov, Phys.Rev.Lett.\textbf{85}, 1158(2000).}
\item {[61]}{A.Albrecht and C.Skordis, Phys.Rev.Lett.\textbf{84},2076(1999)..}
\item {[62]}{C.M.Chen, P.M.Ho, I.P.Neupane and J.E.Wang, JHEP\textbf{0307},017(2003).}
\item {[63]}{C.M.Chen, P.M.Ho, I.P.Neupane, N.Ohta and J.E.Wang, JHEP\textbf{0310}, 058(2003).}
\item {[64]}{I.P.Neupane and D.L.Wiltshire, Phys.Rev.D\textbf{72}, 083509(2005).}
\item {[65]}{S.Tsujikawa, Phys.Rev.D\textbf{73}, 103504(2006).}
\item {[66]}{Y.G.Gong, A.Z.Wang and Y.Z.Zhang, Phys.Lett. B\textbf{636}, 286(2006).}
\item {[67]}{C.Rubano and J.D.Barrow, Phys.Rev.D\textbf{64}, 127301(2001).}
\end{description}}

\end{document}